\documentclass{iaus}
\usepackage{graphicx}

\title[The timescales of chemical enrichment in the Galaxy] 
{The timescales of chemical enrichment in the Galaxy}

\author[A.Pipino \& F.Matteucci] 
{Antonio Pipino$^1$
\and Francesca Matteucci$^2$}

\affiliation{$^1$Physics \& Astronomy, University of Southern California, Los Angeles 90089-0484, USA\\ email: {\tt pipino@usc.edu} \\[\affilskip]
$^2$Dipartimento di Astronomia, Universita'di Trieste, via GB Tiepolo 11, 34100 Trieste, Italy}

\pubyear{2008}
\volume{258} 
\jname{The Ages of Stars}
\editors{eds.}
\begin{document}

\maketitle

\begin{abstract}
The time-scales of chemical enrichment are fundamental to understand the
evolution of abundances and abundance ratios in galaxies. In particular,
the time-scales for the enrichment by SNe II and SNe Ia are crucial in
interpreting the evolution of abundance ratios such as [$\alpha$/Fe]. In fact, the
$\alpha$-elements are produced mainly by SNe II on time-scales of the order of 3
to 30 Myr, whereas the Fe is mainly produced by SNe Ia on a larger range of
time-scales, going from 30 Myr to a Hubble time. This produces differences in
the [$\alpha$/Fe] ratios at high and low redshift and it is known as "time-delay"
model. In this talk we review the most common progenitor models for SNe Ia
and the derived rates together with the effect of the star formation history
on the [$\alpha$/Fe] versus [Fe/H] diagram in the Galaxy. From these diagrams we
can derive the timescale for the formation of the inner halo (roughly 2 Gyr),
the timescale for the formation of the local disk (roughly 7-8 Gyr) as well
the time-scales for the formation of the whole disk. These are functions of
the galactocentric distance and vary from 2-3 Gyr in the inner disk up to a
Hubble time in the outer disk (inside-out formation). Finally, the timescale
for the formation of the bulge is found to be no longer than 0.3 Gyr, similar
to the timescale for the formation of larger spheroids such as elliptical
galaxies.
We show the time-delay model applied to galaxies of different
morphological type, identified by different star formation histories,
and how it constrains differing galaxy formation models.

\end{abstract}

\firstsection 

\section{How to model galactic chemical evolution: role of SNIa vs. SNII}
\label{sec:1}
Before going into the detailed chemical evolution history of the Milky Way, it is necessary to understand how to model, in general, galactic chemical evolution.
The basic ingredients to build a model of galactic chemical evolution 
can be summarized as: i) Initial conditions; ii) Stellar birthrate function (the rate at which stars are formed from the gas and their mass spectrum); iii) Stellar yields (how elements are produced in stars and restored into the interstellar medium); iv) Gas flows (infall, outflow, radial flow). We refer the reader to Matteucci (2008) for a thorough review of their relative roles.
Here we just briefly focus on some properties of both Type II and Ia SNe and recall the mass range of their progenitors:

\subsubsection{Massive stars ($8 < M/M_{\odot} \le 40$)}
In the mass range 10-40 $M_{\odot}$,
available calculations are from Woosley \& Weaver (1995), 
Thielemann et al. (1996),
Meynet \& Maeder (2002), Nomoto et al. (2006), among others. 
These stars end their life as Type II SNe and explode by  core-collapse; 
they  produce  mainly $\alpha$-elements 
(O, Ne, Mg, Si, S, Ca), some Fe-peak elements,  
s-process elements ($A< 90$)
and r-process elements. Lifetimes are below 30 Myr.

\subsubsection{Type Ia SN progenitors}
There is a general consensus about the fact that SNeIa originate from C-deflagration in C-O white dwarfs (WD) 
in binary systems, but several evolutionary paths can lead to such an event.
The C-deflagration produces $\sim 0.6-0.7 M_{\odot}$ of Fe plus traces of other elements from C to Si, as observed in the spectra of Type Ia SNe. Two main evolutionary scenarios for the progenitors of Type Ia SNe have been proposed:

\emph{Single Degenerate (SD) scenario}: the
classical scenario of Whelan and Iben (1973), recently revised by Han \& Podsiadlowsky (2004),  namely C-deflagration in
a C-O WD reaching the Chandrasekhar mass $M_{Ch}\sim 1.44 M_{\odot}$  after accreting material from a red giant
companion. One of the limitations of this scenario is that the accretion rate should be defined in a quite narrow range of values.

The clock to the explosion is given by the lifetime of the secondary star in the binary system, where the WD is the primary (the originally more massive one). Therefore, the largest mass for a secondary is $8 M_{\odot}$, which is the maximum mass for the formation of a C-O WD. As a consequence, the minimum timescale for the occurrence of Type Ia SNe is $\sim 30$ Myr (i.e. the lifetime of a $8M_{\odot}$) after the beginning of star formation. Recent observations in radio-galaxies by Mannucci et al. (2005, 2006) 
seem to confirm the existence of such prompt Type Ia SNe.
The minimum mass for the secondary is $0.8 M_{\odot}$, which is the star with lifetime equal to the age of the Universe. Stars with masses below this limit are obviously not considered.

\emph{Double Degenerate (DD) scenario}:
the merging of two C-O white dwarfs, due to loss of angular momentum caused
by gravitational wave radiation,
which explode by C-deflagration when $M_{Ch}$ is reached (Iben
and Tutukov 1984). In this scenario, the two C-O WDs should be of $\sim 0.7 M_{\odot}$ in order to give rise to a Chandrasekhar mass after they merge, therefore their progenitors should be in the range (5-8)$M_{\odot}$. The clock to the explosion here is given by the lifetime of the secondary star plus the gravitational time delay which depends on the original separation of the two WDs. The minimum timescale for the appearance of the first Type Ia SNe in this scenario is 
one million years more than in the SD scenario (e.g. Greggio 2005 and references therein). At the same time, the maximum gravitational time delay can be as long as more than a Hubble time.

A way of defining the typical Type Ia SN timescale is to assume it as the
time when the maximum in the Type Ia SN rate is reached (Matteucci \& Recchi, 2001). \emph{ This timescale varies according to the chosen progenitor model and to the assumed star formation history, which varies from galaxy to galaxy.}
For the solar vicinity, this timescale  is at least 1 Gyr,
if the SD scenario is assumed, whereas for elliptical galaxies, where the stars formed much more quickly, this timescale is only 0.5 Gyr (Matteucci \& Greggio, 1986; Matteucci \& Recchi 2001).

\section{Chemical evolution time-scales and the formation of the Milky Way}

The kinematical and chemical properties of the different Galactic stellar populations can be interpreted in terms of 
the Galaxy formation mechanism. Here we focus on the results of
the two-infall model of Chiappini, Matteucci \& Gratton (1997) to which we refer the reader for details. Such a scenario predicts two 
main episodes of gas accretion: during the first one, the halo the bulge and most of the thick disk formed, while the second gave rise to the thin disk. 
We first witness the sequence of the formation of the stellar halo, in particular the inner halo, following a monolithic-like collapse of gas (first infall episode) but with a longer timescale than originally suggested by Eggen et al. (1962): here the time scale is 0.8-1.0 Gyr. During the halo formation also the bulge is formed on a very short timescale in the range 0.1-0.5 Gyr (see below).  During this phase also the thick disk assembles or at least part of it, since part of the thick disk, like the outer halo, could have been accreted. Then the thin disk formation, namely the assembly of the innermost disk regions just around the bulge, begins; this is due to the second infall episode. The thin-disk assembles inside-out, in the sense that the outermost regions take a much longer time to form. It is clear that the early phases of the halo and bulge formation are dominated by Type II SNe (and also  by Type Ib/c SNe) producing mostly $\alpha$-elements such as O and Mg and part of Fe. On the other hand, Type Ia SNe start to be non negligible only after 1Gyr and they pollute the gas during the thick and thin disk phases. 
\begin{figure}
\includegraphics[width=12cm,height=6cm]{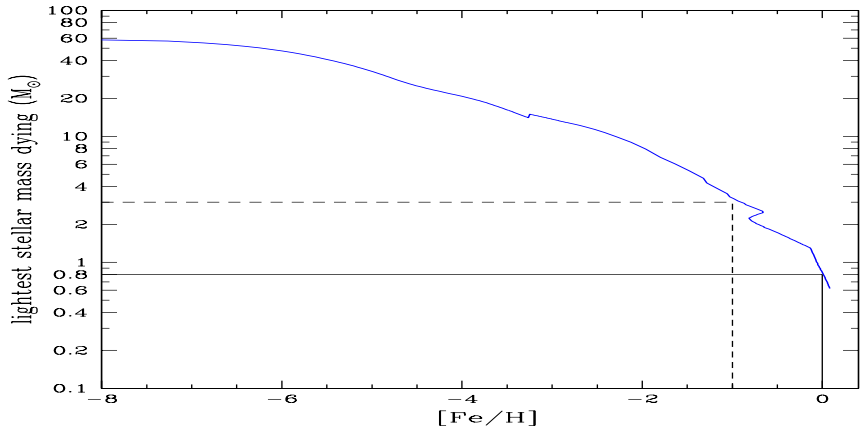}
\hfill
\caption{In this figure we show the smallest stellar mass which dies at any given [Fe/H] achieved by the ISM as a consequence of chemical evolution (see text).}
\label{fig16}
\end{figure}

More in detail, the main assumptions of 
the two-infall model are: the IMF is that of Scalo (1986) normalized over a mass range of 
0.1-100$M_{\odot}$. The infall law gives the rate at which the total surface mass density changes because of the infalling gas. 
It is normalised to reproduce the total present 
time surface mass density in the solar 
vicinity ($\sigma_{tot}$= 51 $\pm$ 6 $M_{\odot} \; pc^{-2}$ , see Boissier \& 
Prantzos 1999), 1 Gyr
is the time for the maximum infall on the thin  disk, 0.8 Gyr
is the e-folding time for the formation of the halo thick-disk (which means a total duration of 2 Gyr for the complete halo-thick disk formation) and $\tau_{D}(r)$
is the timescale for the formation of the thin disk and it is a function of 
the galactocentric distance (formation inside-out, Matteucci \& 
Fran\c cois 1989; Chiappini et al. 2001).
In particular, it is assumed  that $\tau_{D}=1.033 r (Kpc) - 1.267 \,\, (Gyr)$
where $r$ is the galactocentric distance.

The SFR is the Kennicutt law with a dependence on the surface gas 
density and also on the total surface mass density
(see Dopita \& Ryder 1994). The exponent of the surface gas density is set equal to 1.5, similar to what suggested by Kennicutt (1998a).
These choices for the parameters allow the model to fit very well the observational constraints, in particular in the solar vicinity.
We recall that below a critical threshold for the surface gas density 
($7M_{\odot}pc^{-2}$ for the thin disk and $4M_{\odot}pc^{-2}$ for the halo phase)
we assume that the star formation is halted. 
The existence of a threshold for the star formation has been suggested by Kennicutt (1998a,b).

\subsection{The chemical enrichment history of the solar vicinity}

In Fig.~\ref{fig16} we show the smallest mass dying at any cosmic time corresponding to a given predicted 
abundance of [Fe/H] in the ISM. This is because there is an age-metallicity relation and the 
[Fe/H] abundance increases with time. Thus, it is clear that in the early phases of the halo only massive stars are dying and contributing to the chemical enrichment process. Clearly this graph depends upon the assumed stellar lifetimes and upon the age-[Fe/H] relation.
It is worth noting that the Fe production from Type Ia SNe appears before the gas has reached [Fe/H] =-1.0, therefore during the halo and thick disk phase. This clearly depends upon the assumed Type Ia SN progenitors (in this case the single degenerate model).
\begin{figure}
\includegraphics[width=12cm,height=8cm]{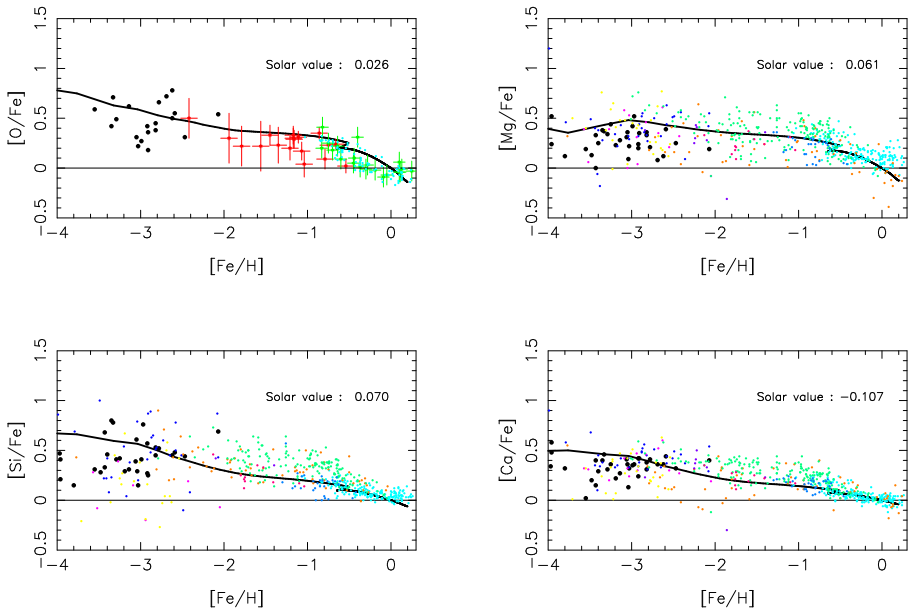}
\caption{Predicted and observed [$\alpha$/Fe] vs. [Fe/H] in the solar neighborhood.
The models and the data are from Fran\c cois et al. (2004).
The models are normalized to the predicted solar abundances. The predicted abundance ratios at the time of the Sun formation (Solar value) are shown in each panel and indicate a good fit (all the values are close to zero).}
\label{fig18}
\end{figure}

\subsubsection{The time-delay model}
\begin{figure}
\centering
\includegraphics[width=10cm,height=8cm]{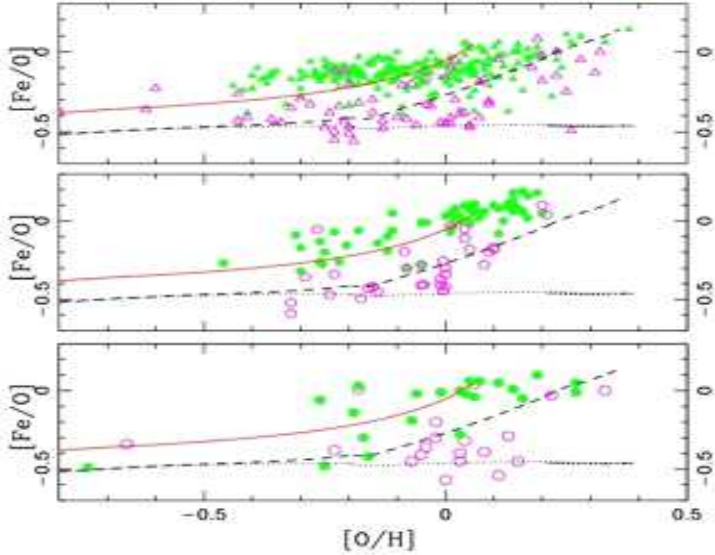}
\caption{Thick vs thin disk [O/Fe]-[Fe/H] diagrams. Theoretical predictions for the thick
and the thin disk are shown by dashed and solid lines, respectively. The dotted line
is a model for the thick disk without pollution by SNIa. Data for the thick
and the thin disk are represented by open and full symbols, respectively. For details
refer to Chiappini (2008).}
\label{fig13}
\end{figure}

The time-delay refers to the delay with which Fe is ejected into the ISM by SNe Ia relative to the fast production of $\alpha$-elements by core-collapse SNe. Tinsley (1979) first suggested that this time delay would have produced a typical signature in the [$\alpha$/Fe] vs. [Fe/H] diagram. Matteucci \& Greggio (1986) included for the first time the Type Ia SN rate formulated by Greggio \& Renzini (1983) in a detailed numerical model for the chemical evolution of the Milky Way. The effect of the delayed Fe production is to create an overabundance 
of O relative to Fe ([O/Fe]$>$ 0) at low [Fe/H] values, and a continuous decline  of the [O/Fe] ratio until the solar value ([O/Fe]$_{\odot}=0.0$) is reached for [Fe/H]$>-1.0$ dex. This is what is observed and indicates that during the halo phase the [O/Fe] ratio is due only to the production of O and Fe by SNe II. However, since the bulk of Fe is produced by Type Ia SNe, when these latter start to be important then the [O/Fe] ratio begins to decline. This effect was predicted by Matteucci \& Greggio (1986) to occur also for other $\alpha$-elements (e.g. Mg, Si). At the present time, a great amount of stellar abundances is available and the trend of the $\alpha$-elements has been since long confirmed. 
In Fig.~\ref{fig18} shows one among our most recent models applied to the latest compilation of data (Fran\c ois et al. 2004).
A good fit of the [O/Fe] ratio as a function of [Fe/H] is obtained only if the $\alpha$-elements are mainly produced by Type II SNe and the Fe by Type Ia SNe. If one assumes that only SNe Ia produce Fe as well as if one assumes that only Type II SNe produce Fe, the agreement with observations is lost. Therefore, the conclusion is that both Types of SNe should produce Fe in the proportions of 1/3 for Type II SNe and 2/3 for Type Ia SNe. 
The model in Fig. ~\ref{fig18} does not distinguish between the thin and the thick disk. Recently, Chiappini (2008) presented a model where the evolutions of the thin and thick disks were considered separately. The main hypothesis was that the thick disk formed and evolved faster than the thin disk (see Fig.~\ref{fig13}), and the conclusion was that a good agreement with the observed abundances in thick disk stars can be obtained if the thick disk assembled by gas accretion on a timescale no longer than 2 Gyr. This conclusion favors the formation of the thick disk stars in situ instead than by accretion of extant stellar satellites. 

\subsubsection{The G-dwarf metallicity distribution and constraints on the thin disk formation}
\begin{figure}
\includegraphics[width=10cm,height=6cm]{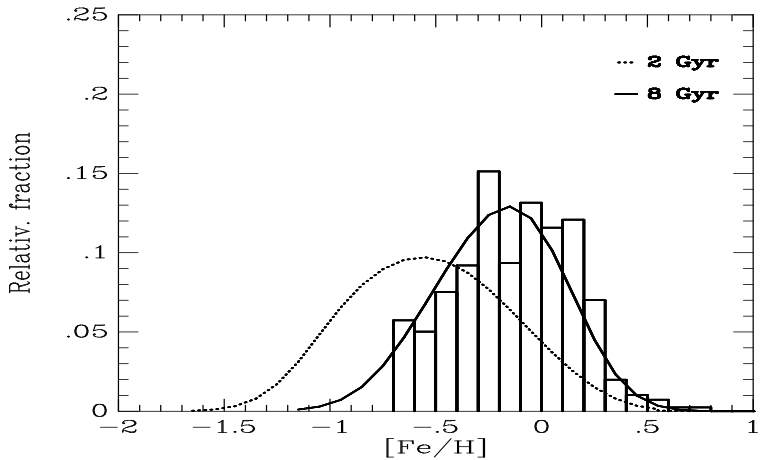}
\caption{The figure is from Kotoneva et al. (2002) and shows the comparison between a sample of K-dwarfs and model predictions in the solar neighborhood.
The dotted curve refers to the two-infall model with a timescale $\tau=2$ Gyr, whereas the continuous line refers to $\tau= 8$ Gyr.}
\label{fig24}
\end{figure}

The G-dwarf metallicity distribution is quite an important constraint for the chemical evolution of the solar vicinity: it is the fossil record of the star formation history of the thin disk. 
Originally, there was the ``G-dwarf problem'' which means that the Simple Model of galactic chemical evolution could not reproduce the distribution of the G-dwarfs. 
It has been since long demonstrated that relaxing the closed-box assumption and allowing for the solar region to form gradually by accretion of gas can solve the problem (Tinsley, 1980). Assuming that the disk forms from pre-enriched gas can also solve the problem but still the gas infall is necessary to have a realistic picture of the 
disk formation. 
The two-infall model can reproduce very well the G-dwarf distribution and also that of K-dwarfs (see Fig.~\ref{fig24}),  as long as a timescale for the formation of the disk in the solar vicinity of 7-8 Gyr is assumed. This conclusion is shared by other authors (Alib\'es et al. 2001; Prantzos \& Boissier 1999)

\subsection{The Galactic disk}
The chemical abundances measured along the disk of the Galaxy suggest that the metal content decreases from the innermost to the outermost regions, in other words there is a negative gradient in metals.

\begin{figure}
\includegraphics[width=12cm,height=8cm]{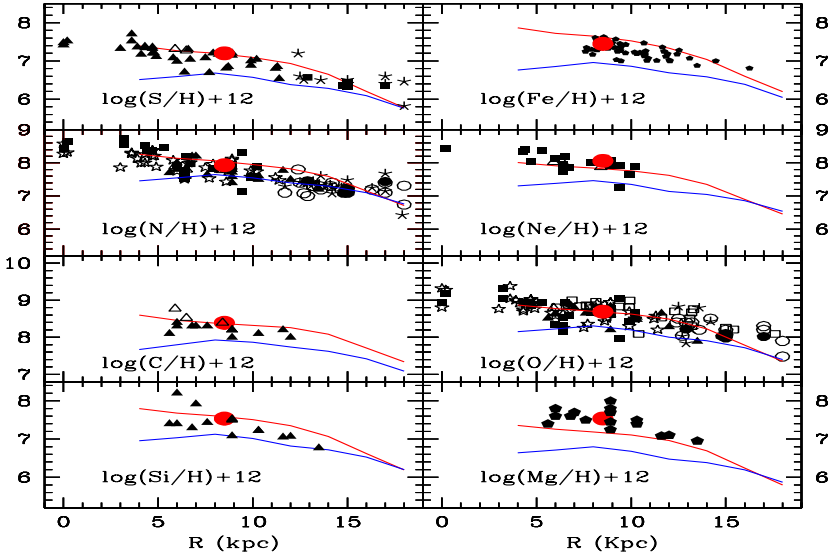}
\hfill
\caption{Spatial and temporal behaviour of abundance 
gradients along the Galactic disk as predicted by the best model of Chiappini et al. (2001). The upper lines in each panel represent the present time gradient, whereas the lower ones represent the gradient a few Gyr ago (see Chiappini et al. 2001).} 
\label{fig30}
\end{figure}

In Fig.~\ref{fig30} we show theoretical predictions of abundance gradients along the disk of the Milky Way compared with data from HII regions, B stars and PNe. The adopted model is from Chiappini et al. (2001) and is based on an 
inside-out formation of the thin disk. The model does not allow for exchange of gas between different regions of the disk. The disk is, in fact, divided in several concentric shells 2 Kpc wide with no interaction between them.

The reason for the steepening is that in 
the model of Chiappini et al. there is included a threshold density for SF, which 
induces the SF to stop when the density decreases below the threshold. 
This effect is particularly strong in the external 
regions of the Galactic disk, thus
contributing to a slower evolution and therefore to a steepening of the 
gradients with time.
Other authors (e.g. Boissier \& Prantzos, 1999) found a flattening of the 
gradient with time in absence of a star formation threshold.

\subsection{The Galactic bulge}

In the context of chemical evolution, the Galactic bulge was first
modeled by Matteucci \& Brocato (1990) who predicted that the
[$\alpha$/Fe] ratio for some elements (O, Si and Mg) should be 
super-solar over almost the whole metallicity range, in analogy with
the halo stars, as a consequence of assuming a fast bulge evolution
which involved rapid gas enrichment in Fe mainly by Type II SNe. 
Recent data  concerning medium- and high-resolution spectroscopy of bulge stars
(Rich \& McWilliam, 2000; Fulbright et al., 2006, 2007;
Zoccali et al., 2006; Lecureur et al. 2007) and more recent models (e.g. Ballero et al. 2007) have confirmed the previous predictions and suggested that the bulge must have formed on a time scale of $\sim 0.3$ Gyr and in any case no longer than 0.5 Gyr. This is in agreement with 
Elmegreen (1999) and Ferreras et al. (2003). A similar model, although updated with the inclusion of the development of a galactic wind and more recent stellar yields,  has been presented by Pipino et al. (2008b): it shows how a model with intense star formation and rapid assembly of gas
can best reproduce the most recent accurate data on abundance ratios (both in single stars and in the integrated
spectra) as well as the metallicity distribution .

\section{From the Bulge to external galaxies: Ellipticals}

\begin{figure}
\includegraphics[width=12cm,height=8cm]{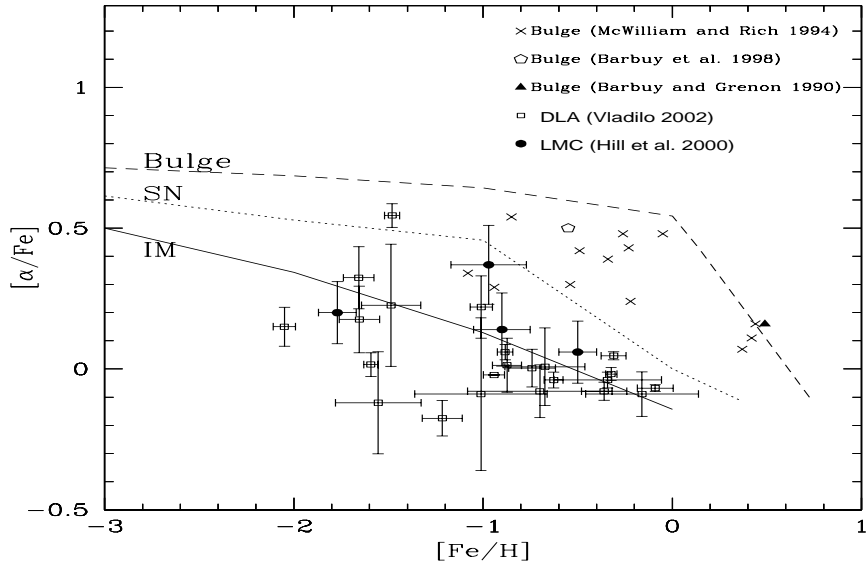}
\hfill
\caption{The predicted [$\alpha$/Fe] vs. [Fe/H] relations for the Galactic bulge (upper curve), the solar vicinity (median curve) and irregular galaxies (low curve). Data for the bulge are reported for comparison. Data for the LMC and DLA systems are also shown for comparison, indicating that DLAs are probably irregular galaxies. Figure and references are from Matteucci 
(2003).}\label{fig33}
\end{figure}

In Fig.~\ref{fig33} we present the predictions by Matteucci (2003) of the 
[$\alpha$/Fe] ratios as functions of [Fe/H] in galaxies of different morphological type. In particular, for the Galactic bulge or an elliptical galaxy of the same mass, for the solar vicinity region and for an irregular Magellanic galaxy (LMC and SMC).
The underlying assumption is that different objects undergo different  histories of star formation, being very fast in the spheroids (bulges and ellipticals), moderate in spiral disks and slow and perhaps gasping in irregular gas rich galaxies. The effect of different star formation histories is evident in Fig.~\ref{fig33} where the predicted  [$\alpha$/Fe] ratios in the bulge and ellipticals remain high and almost constant for a large interval of [Fe/H]. This is due to the fact that, since star formation is very intense, the bulge reaches very soon a solar metallicity thanks only to the SNe II; then, when SNe Ia start exploding and ejecting Fe into the ISM, 
the change in the slope occurs at larger [Fe/H] than in the solar vicinity.
In the extreme case of irregular galaxies the situation is opposite: here the star formation is slow and when the SNe Ia start exploding the gas is still very metal poor. 
This scheme is quite useful since it can be used to identify galaxies only by looking at their abundance ratios. 


The above mentioned models for the bulge can be extended to study elliptical galaxies.
In fact, similarly to bulges, elliptical galaxies show over-solar [$\alpha$/Fe] ratios.
Moreover, these ratios are found to increase with galactic mass.
In the framework of the time-delay model, we can explain such a behaviour
as an increase of the SF efficiency as a function
of galactic mass (Matteucci 1994, Pipino \& Matteucci, 2004).
To explain the higher star formation efficiency in the most massive galaxies in the framework
of the \emph{monolithic collapse},
Pipino et al. (2008c) appeal to massive black holes-triggered SF:
a short ($10^6-10^7$ yr) super-Eddington phase can 
provide the accelerated triggering of associated star formation.
According to Pipino et al. (2008c) models, the galaxy is fully assembled on a time-scale of 0.3-0.5 Gyr.

\begin{figure}
\includegraphics[width=12cm]{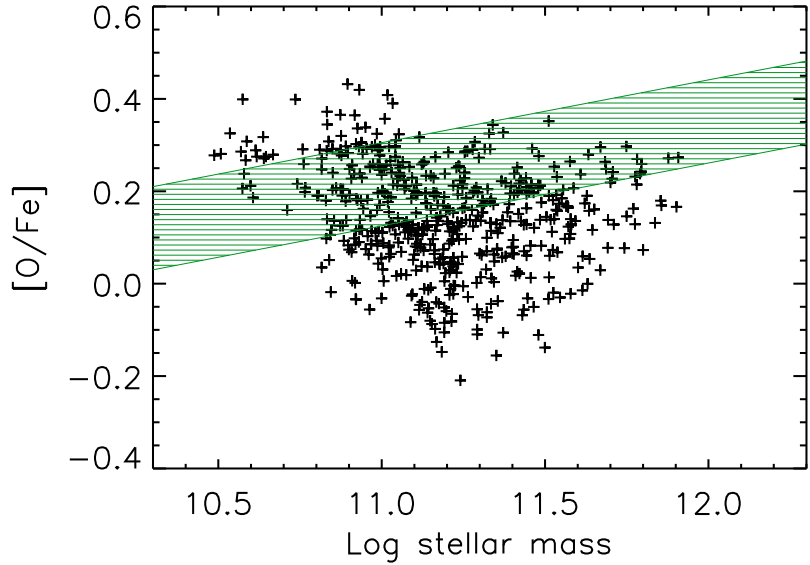}
\caption{The $\alpha/Fe$-mass and relations as predicted by GalICS (Pipino et al. 2008a) for the whole 
sample of ellipticals (black points). 
The thick solid lines
encompass the 1$\sigma$-region (hatched area) around the mean trend reported by Thomas et al. (2008, in prep.).
\label{mfmr}}
\end{figure}

On the other hand, when the same detailed treatment for the chemical evolution is implemented in
a semi-analytic model for galaxy formation based on the \emph{hierarchical clustering scenario} (e.g. White \& Rees, 1978), 
it does not produce the observed mass-[$\alpha$/Fe] relation (Pipino et al. 2008a). 
In particular, the slope is too shallow and the scatter too large (see Fig.~\ref{mfmr}), 
in particular in the low and intermediate mass range. The model shows significant improvement at the highest masses and 
velocity dispersions, where the predicted [$\alpha$/Fe] ratios are now marginally consistent with observed values. 
Moreover, an excess of low-mass ellipticals
with too high a [$\alpha$/Fe] ratio is predicted.

A thorough exploration of the parameter space shows
that the failure in reproducing the mass- and $\sigma$-[$\alpha$/Fe] relations can partly be attributed to the way in which star formation and feedback are currently modelled. The merger process is responsible for a part of the scatter. We suggest that the next 
generation of semi-analytical model should feature feedback (either stellar of from AGN) mechanisms linked to single galaxies and not only to the halo, especially in the low and intermediate mass range.
The scatter is also intrinsic to the merger history,
thus calling for further modification
of the baryons behaviour with respect to the CDM.
In other words we envisage a lack of a self-regulating
mechanisms which acts on a galactic scale and counterbalances
to some extent the random nature of the merger trees.

\section{Conclusions}

From the discussions of the previous sections we can extract some important 
conclusions on the formation and evolution of the Milky Way, derived from 
chemical abundances.
In particular, the inner halo formed on a timescale of 1-2 Gyr at maximum, the outer halo 
formed on longer time-scales perhaps from accretion of satellites or gas. 
The disk at the solar ring formed on a timescale not shorter than 7 Gyr.
The whole disk formed inside out with time-scales of the order of 2 Gyr 
or less in the inner regions and 10 Gyr or more in the outermost regions.
The bulge is very old and formed very quickly on a timescale smaller 
than even the inner halo and  not larger than 0.5 Gyr.
The abundance gradients arise naturally from the assumption of the 
inside-out formation of  the disk. A threshold density for the star formation helps in steepening the gradients in the outer disk regions.
The IMF seems to be different in the bulge and the disk, being flatter 
in the bulge, although more abundance data are necessary before drawing firm 
conclusions.
Elliptical galaxies and bulges share similar chemical properties
(at a given mass), therefore a very short and intense formation
is predicted for them.
While this can be accommodated in the \emph{monolithic framework} by means
of, e.g., AGN-triggered star formation, it is still hard
for models based on \emph{hierarchical clustering} to account for the observed
chemical properties.





\end{document}